\documentclass[amsmath,amssymb,amsbsy,pre,nofootinbib,twocolumn,tightenlines,superscriptaddress,floatfix]{revtex4-2}
\usepackage{amsfonts} 
\usepackage{amsmath} 
\usepackage{blindtext}
\usepackage{lineno}
\usepackage{xfrac,mathtools}
\usepackage{cancel}
\usepackage{bm}
\usepackage{booktabs}
\usepackage[usenames, dvipsnames]{color}
\usepackage{dcolumn}
\usepackage{epic} 
\usepackage{epsfig}
\usepackage{esdiff}
\usepackage{nicefrac}
\usepackage{graphicx}
\usepackage{xcolor}
\usepackage{soul}
\usepackage[breaklinks,colorlinks = true,linkcolor = blue,urlcolor=blue,citecolor=red]{hyperref}

\begin{document}
	\title{Dynamical slowdown, bottlenecks, and multiscaling in Voigt-regularised turbulence}
	\author{Anikat Kankaria$^\clubsuit$}
	\email{anikat.kankaria@icts.res.in}
	\affiliation{International Centre for Theoretical Sciences, Tata Institute of Fundamental Research, Bengaluru 560089, India}
	\author{Bikram Pal$^\clubsuit$}
	\email{bikram25052005@gmail.com}
	\affiliation{Department of Physical Sciences, Indian Institute of Science Education and Research Kolkata, Mohanpur 741246, India}
	\author{Edriss S. Titi}
	\email{edriss.titi@maths.cam.ac.uk}
	\affiliation{Department of Applied Mathematics and Theoretical Physics, University of Cambridge, Cambridge CB3 0WA, UK}
	\affiliation{Department of Mathematics, Texas A \& M University, College Station, TX 77843-3368, USA}
	\affiliation{Department of Computer Science and Applied Mathematics, Weizmann Institute of Science, Rehovot 76100, Israel}
	\author{Samriddhi Sankar Ray}
	\email{samriddhisankarray@gmail.com}
	\affiliation{International Centre for Theoretical Sciences, Tata Institute of Fundamental Research, Bengaluru 560089, India}

	\begin{abstract} We investigate bottleneck formation in turbulence
		using the Voigt--regularised SABRA shell model and direct
		numerical simulations of the corresponding
		Navier--Stokes--Voigt equations. The Voigt regularisation
		introduces a scale-dependent slowdown of nonlinear interactions
		without directly enhancing dissipation, providing a natural
		setting in which to study the interplay between nonlinear
		transfer and equilibrium-like behaviour.  We find three
		distinct spectral regimes: a Kolmogorov inertial range at
		$k<k_{\rm I}$, an intermediate equilibrium-like range
		associated with partial thermalisation for $k_{\rm I}<k<k_{\rm
			II}$, and a high-wavenumber equilibrium regime for $k>k_{\rm
			II}$, where the Voigt contribution dominates the conserved
		invariant. The crossover to the high-wavenumber regime occurs
		at $k_{\rm II}\sim\alpha^{-1}$, while $k_{\rm I}$ marks the
		onset of equilibrium-like behaviour. Equal-time and dynamical
		statistics reveal a progressive suppression of intermittency
		and a tendency towards Gaussian behaviour at small scales,
		together with a transition from dynamic multiscaling in the
		turbulent regime to simple scaling in the equilibrium ranges.
		The shell model resolves these three regimes over a broad range
		of scales, while Direct Numerical Simulations of the
		corresponding Navier--Stokes--Voigt equations reproduces the
		same qualitative trends, including bottleneck formation,
		delayed cascade completion, reduced intermittency, and a
		tendency towards Gaussian statistics at small scales.  Our
		results suggest that bottleneck formation might be associated
		with scale-dependent dynamical slowdown and incipient
		thermalisation, rather than being purely dissipative in origin.
		Furthermore, we provide strong evidence that,
		in the limiting regime where the regularization parameter
		$\alpha$ is much smaller than the dissipation length scale, the
		Voigt model reproduces the same inertial-range turbulent regime
		and turbulence statistics as the Navier–Stokes equations. This
		provides further evidence that the Voigt model constitutes an
		excellent practical approximation to the Navier–Stokes
		equations for small values of $\alpha$.
	\end{abstract}
	
	\maketitle
	
	\begingroup
	\renewcommand{\thefootnote}{$\clubsuit$}
	\footnotetext{These authors contributed equally to this work.}
	\endgroup

	\section{Introduction}
	
	Understanding the origin of bottleneck phenomena~\cite{Falkovich1994,Lohse1995,Ishihara2009} in turbulent flows remains an
	open problem in fully developed turbulence~\cite{Frisch95}. Bottlenecks appear as pronounced
	accumulations of energy near the high-wavenumber end of the inertial range and
	have been observed in experiments~\cite{Pak1991ExpBottleneck,She1993ExpBottleneck,Saddoughi1994ExpBottleneck} and 
	numerical simulations~\cite{She1993,Yeung1997,Gotoh2002,Kaneda2003,Dobler2003,Kurien2004,Verma2007,Mininni2008,Donzis2010}.
	Although their existence is well established, the physical mechanisms
	responsible for their formation remain the subject of some debate.
	
	Early explanations~\cite{Falkovich1994,Lohse1995} emphasised the progressive inefficiency of the
	forward energy cascade near the dissipative scales. In this picture,
	the reduction of nonlinear transfer leads to a local accumulation of
	energy and hence to the characteristic bump in the spectrum.
	Such interpretations invoke a dynamical pile-up of energy rather than
	an equilibrium state and have proved successful in explaining many
	observed features of the bottleneck phenomenon~\cite{Kurien2004,Verma2007,Donzis2010}.
	
	A complementary viewpoint relates bottlenecks to partial thermalisation induced 
	by a hyperviscous term in the Navier-Stokes equation~\cite{Frisch2008,Frisch2013,Banerjee2014}.  In
	systems where the forward cascade becomes sufficiently inefficient before
	dissipation fully dominates, the dynamics may explore equilibrium-like states
	over a restricted range of scales. Such behaviour is well known in spectrally
	truncated Euler dynamics~\cite{Cichowlas2005,Bos2006,Ray2015,Venkataraman2017,Dileoni2017,Murugan2020,Murugan2023}, where the absence of dissipation leads to absolute
	equilibrium and Gaussian statistics at large wavenumbers~\cite{Krstulovic2009,Murugan2021,Rajarshi2026}. Whether similar
	mechanisms operate, even locally and approximately, in fully developed
	turbulence remains an important open question.
	
	Regularised models of turbulence provide a natural setting in which to
	investigate how modifications of the cascade dynamics influence bottleneck
	formation and the possible emergence of equilibrium-like behaviour. The
	$\alpha$--Voigt regularisation was introduced by Oskolkov
	\cite{Oskolkov1973,Oskolkov1980,Oskolkov1985} as a regularisation of the
	Navier--Stokes equations. Its connection to turbulence modelling was
	subsequently developed by Cao, Lunasin, and Titi~\cite{Cao2006}, who
	identified the corresponding simplified Bardina/Voigt formulation as a
	possible subgrid-scale model for turbulence. The statistical theory of the
	Voigt equations was subsequently developed through the construction of
	invariant probability measures for the three-dimensional
	Navier--Stokes--Voigt equations and the analysis of their convergence to
	invariant measures of the Navier--Stokes equations as
	$\alpha\to0$~\cite{RamosTiti2010}. This turbulence perspective was further pursued numerically in the
	Voigt--SABRA shell model by Levant, Ramos, and Titi~\cite{Levant2010},
	who reported distinct spectral regimes and a reduction of intermittency
	with increasing Voigt regularisation.  
	Subsequent work has explored the Voigt equations from complementary
	perspectives, including mathematical analysis of well-posedness and
	statistical solutions~\cite{Larios2010,Berselli2012,Larios2014} and numerical
	studies of self-truncation, thermalisation, and turbulent spectra
	\cite{DiMolfetta2015,Larios2018}. Unlike hyperviscosity or ad hoc spectral
	filtering, the Voigt regularisation does not directly enhance dissipation.
	Instead, it modifies the nonlinear timescale itself, providing a controlled
	setting in which to study the consequences of dynamical slowdown for
	turbulent energy transfer.
	
	From this perspective, the Voigt model offers a natural setting in which to
	investigate bottleneck formation. As the nonlinear transfer time increases
	with wavenumber, the forward cascade becomes progressively less efficient,
	allowing energy to accumulate locally in scale space. Bottlenecks may then be
	viewed as signatures of incipient thermalisation generated by a breakdown of
	cascade dynamics rather than as purely dissipative phenomena.
	
	To explore this possibility, we employ the SABRA shell model~\cite{Lvov1998}
	of the Voigt-regularised Navier--Stokes equations~\cite{Larios2018} together with direct
	numerical simulations of the corresponding hydrodynamic system. Shell models
	retain the essential ingredients of turbulence~\cite{Frisch95,Biferale2003,Pandit2009,AlexakisBiferale2018}, including nonlinear transfer,
	cascade dynamics, and conservation laws, while allowing substantially larger
	scale separations and improved statistical convergence than DNS. This makes
	them particularly well suited for studying the interplay between turbulent and
	equilibrium dynamics~\cite{Lvov2002,Tom2017} over a wide range of scales.
	
	Building on the spectral phenomenology reported in the
	Voigt--SABRA shell model~\cite{Levant2010}, we revisit the emergence of
	multiple spectral regimes with substantially greater scale separation
	and provide a systematic characterisation of their spectral,
	statistical, and dynamical properties. The bottleneck is not a single
	spectral feature but is associated with the coexistence of three
	distinct regimes. At the largest scales the dynamics remains turbulent
	and exhibits Kolmogorov-like behaviour. At intermediate scales the
	system enters an equilibrium-like state associated with partial
	thermalisation, while at sufficiently large wavenumbers a second
	equilibrium regime emerges in which the Voigt contribution to the
	conserved invariant becomes dominant.  We show that the crossover to
	the latter regime occurs at a scale $k_{\rm II}\sim\alpha^{-1}$, while
	the onset of the intermediate regime is characterised by a second scale
	$k_{\rm I}$, whose asymptotic behaviour is derived in Sec. III and
	Appendix~\hyperref[ap:derivation]{A}.
	
	To substantiate this interpretation, we examine energy spectra, equal-time
	structure functions, probability distribution functions, and time-dependent
	structure functions. The spectral and statistical properties of the two
	equilibrium regimes are shown to be consistent with equipartition arguments
	based on the conserved Voigt invariant. Furthermore, the temporal dynamics
	undergoes a corresponding transition from turbulent dynamic multiscaling to
	simple scaling governed by a single characteristic relaxation timescale.
	Although the available scale separation in DNS is necessarily more limited,
	the DNS results display the same qualitative trends observed in the shell
	model, including bottleneck formation, delayed cascade completion, reduced
	intermittency, and a tendency towards Gaussian statistics at small scales.
	
	Taken together, these results support a picture in which bottlenecks arise as
	a consequence of scale-dependent dynamical slowdown and incipient
	thermalisation. More broadly, they suggest that the Voigt model provides a
	particularly transparent framework for studying the competition between
	nonequilibrium cascade dynamics and equilibrium behaviour in turbulent flows.
	
	\section{Voigt equations, shell-model formulation, and numerical setup}
	
	We begin with the incompressible Navier--Stokes--Voigt equations, given by~\cite{Oskolkov1973}
	\begin{equation}
		\left\{
		\begin{aligned}
			&\partial_t \left( \mathbf{v} - \alpha^2 \nabla^2 \mathbf{v} \right)
			+ \mathbf{v}\cdot\nabla \mathbf{v}
			= -\nabla p + \nu \nabla^2 \mathbf{v} + \mathbf{f}, \\
			&\nabla\cdot\mathbf{v} = 0,
		\end{aligned}
		\right.
		\label{eq:NSV}
	\end{equation}
	where $\mathbf{v}(\mathbf{x},t)$ is the velocity field, $p$ the pressure,
	$\nu$ the kinematic viscosity, and $\mathbf{f}$ an external forcing acting at
	large scales. The parameter $\alpha$ has dimensions of length and introduces
	a scale-dependent modification of the time derivative. In the limit
	$\alpha\to0$, equation~\eqref{eq:NSV} reduces to the incompressible
	Navier--Stokes equations.
	
	In the inviscid and unforced limit, equation~\eqref{eq:NSV} conserves a quadratic
	invariant that differs from the standard kinetic energy
	\cite{Levant2010}. Throughout this work we employ
	both direct numerical simulations (DNS) of equation~\eqref{eq:NSV} and a
	Voigt-regularised shell-model representation. The DNS allows us to verify that
	the observed phenomenology persists in the underlying hydrodynamic equations,
	while the shell model provides access to substantially larger scale
	separations and improved statistical convergence.
	
	To explore a broader range of scales than is accessible in DNS, we consider
	the Voigt-regularised SABRA shell model. We introduce complex shell variables
	$u_n(t)$ associated with wavenumbers
	\begin{equation}
		k_n = k_0 \lambda^n,
		\qquad n=0,1,2,\ldots,N-1,
	\end{equation}
	where $\lambda>1$ is the shell spacing parameter. The Voigt-regularised
	SABRA shell model is defined by
	\begin{equation}
		\left(1+\alpha^2 k_n^2\right)\dot{u}_n
		=
		N_n(u)-\nu k_n^2 u_n+f_n,
		\label{eq:VoigtSABRA}
	\end{equation}
	where $f_n$ denotes forcing applied at the largest scales and
	$N_n(u)$ is the standard SABRA nonlinear interaction,
	\begin{equation}
		\begin{aligned}
			N_n(u) &= i\Big( a k_{n+1}u_{n+2}u_{n+1}^{*} +b k_n u_{n+1}u_{n-1}^{*} \\ &\qquad -c k_{n-1}u_{n-1}u_{n-2} \Big),
		\end{aligned}
	\end{equation}
	with real coefficients satisfying $a+b+c=0.$
	This condition guarantees conservation of the quadratic invariant in the
	inviscid and unforced limit.
	
	The inviscid, unforced Voigt--SABRA model conserves the quantity
	\begin{equation}
		E_\alpha = \sum_n \left(1+\alpha^2 k_n^2\right)|u_n|^2,
		\label{eq:VoigtEnergy}
	\end{equation}
	which we refer to as the Voigt-modified energy. In contrast, the standard
	shell-model energy
	\begin{equation}
		E = \sum_n |u_n|^2,
	\end{equation}
	is no longer conserved when $\alpha\neq0$.
	
	The corresponding Voigt-modified energy spectrum is defined as
	\begin{equation}
		E_\alpha(k_n) = \frac{1}{2} \left(1+\alpha^2 k_n^2\right) \frac{\langle |u_n|^2\rangle}{k_n},
	\end{equation}
	while the standard energy spectrum is
	\begin{equation}
		E(k_n) = \frac{1}{2} \frac{\langle |u_n|^2\rangle}{k_n}.
	\end{equation}
	
	The Voigt regularisation also modifies the characteristic nonlinear time
	scale. A dimensional estimate gives
	\begin{equation}
		\tau_{\mathrm{nl}}^{-1}(k_n) \sim \frac{k_n |u_n|} {1+\alpha^2 k_n^2},
		\label{eq:nltime}
	\end{equation}
	which introduces an explicit wavenumber dependence into the nonlinear
	dynamics. The consequences of this modified timescale for the spectral,
	statistical, and dynamical properties of the system are examined in the
	following sections.
	
	The Voigt--SABRA shell-model equations were integrated using a fourth-order
	Runge--Kutta (RK4) time-stepping scheme. Simulations were performed for
	Voigt parameters in the range $ 10^{-7} \le \alpha \le 10^{-5}$, together with the
	standard SABRA model ($\alpha=0$), which serves as a reference case
	throughout this work. In the simulations reported here we choose
	$\lambda = 2,\, k_0 = 1/2048$, viscosity $\nu = 10^{-14}$, forcing  $f_n = 0.003\times(1 + i)  \delta_{2,n}$,
	and a total of $N = 48$ shells. 
	
	The Navier--Stokes--Voigt equations were solved using phase-shifted de-aliased
	pseudo-spectral method in a triply periodic domain together with a
	fourth-order Runge--Kutta (RK4) time integration scheme. Simulations were
	performed over the range $0 \le \alpha \le 0.2$. We employ $N^3 = 512^3$ collocation points
	and viscosity $\nu = 5 \times 10^{-4}$, and a large-scale constant energy injection forcing 
	yielding Taylor-scale Reynolds numbers of
	approximately $Re_\lambda \approx 295$ in the $\alpha=0$ case. 
	
	\begin{figure*}
		\includegraphics[width = \linewidth]{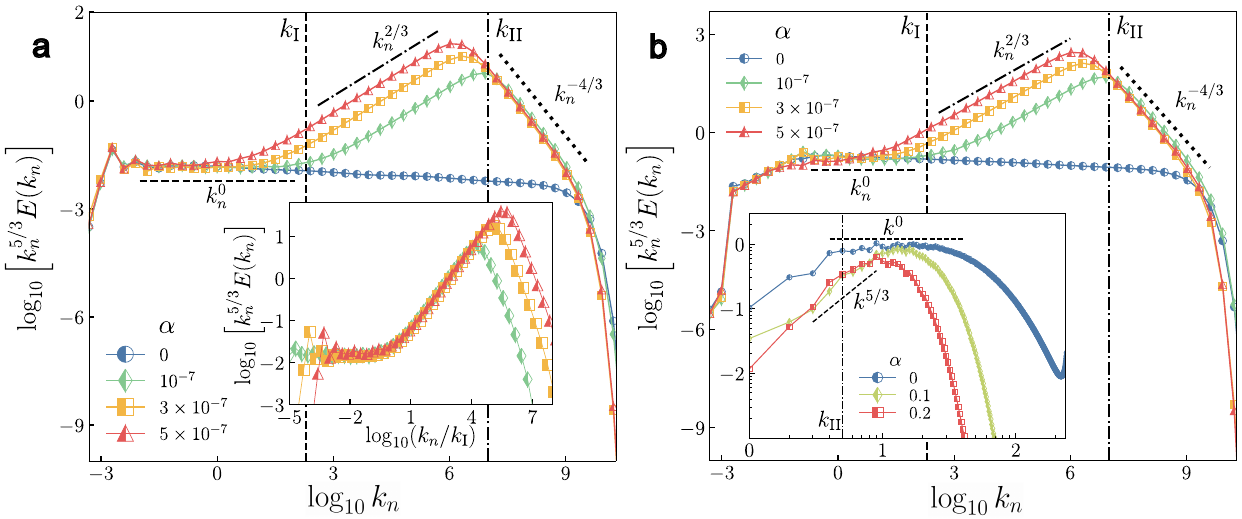}
		\caption{(a) Compensated energy spectra, $k^{5/3}E(k)$, obtained from statistically
			stationary forced Voigt--SABRA simulations for different values of the Voigt
			parameter $\alpha$ (see legend). The vertical dashed and dash-dotted lines
			indicate the crossover wavenumbers $k_{\rm I}$ and $k_{\rm II}$, respectively, for the
			smallest non-zero value of $\alpha$. The inset shows the corresponding compensated
			spectra with the wavenumbers normalised by $k_{\rm I}$.
			(b) Compensated energy spectra from decaying Voigt--SABRA simulations,
			evaluated at the time of maximum enstrophy for each value of $\alpha$ with the pair of 
			vertical dashed and dash-dotted lines indicating $k_{\rm I}$ and $k_{\rm II}$. The
			inset shows the corresponding DNS results for the decaying
			Navier--Stokes--Voigt equations, evaluated at the same stage of the cascade
			evolution. The dash-dotted line is an indication of $k_{\rm II}$ for the largest 
			$\alpha$.}
		\label{fig:CompSpectra}
	\end{figure*}

	\section{Spectral phenomenology and bottleneck formation}
	
	We begin by examining the energy spectrum, since it provides the most direct
	diagnostic of how the modified nonlinear timescale affects energy transfer
	across scale and hence the redistribution of energy across scales. It therefore
	offers a natural framework for investigating bottleneck formation in
	Voigt-regularised turbulence. In this section we focus exclusively on the
	spectral properties of the system and use them to identify the distinct
	dynamical regimes that emerge as the Voigt parameter is increased.
	
	Figure~\ref{fig:CompSpectra} summarises the principal spectral results.
	The main panels show spectra obtained from the Voigt--SABRA shell model,
	while the insets display the corresponding spectra from direct numerical
	simulations (DNS) of the Navier--Stokes--Voigt equations; the $\alpha=0$
	Navier--Stokes solutions are shown as a point of comparison. We begin by
	discussing the shell-model results, where the substantially larger scale
	separation allows the different spectral regimes to be resolved most clearly.
	Indeed, the existence of two distinct crossover scales requires substantial
	scale separation in order for the associated spectral ranges to be identified,
	a condition that is readily achieved in shell models but remains challenging
	in DNS.
	
	Figure~\ref{fig:CompSpectra}(a) shows compensated energy spectra obtained
	from statistically stationary forced simulations of the Voigt--SABRA model.
	For $\alpha=0$, the spectra exhibit an extended plateau consistent with
	Kolmogorov scaling, which terminates in the dissipation range. As
	$\alpha \neq 0$ and increases, however, the inertial-range plateau
	progressively shrinks and a pronounced bottleneck develops at intermediate
	and larger wavenumbers before eventually decreasing again.
	
	The bottleneck does not appear as a simple crossover between inertial and
	dissipative scales. Instead, the shell-model spectra reveal three distinct
	scaling ranges separated by two characteristic wavenumbers, denoted by
	$k_{\rm I}$ and $k_{\rm II}$. At low wavenumbers, the spectra remain essentially
	indistinguishable from those of the standard SABRA model and follow the
	familiar Kolmogorov scaling law $E(k)\sim k^{-5/3}$.
	
	Beyond the first crossover scale $k_{\rm I}$, the spectra enter a second regime
	characterised by a considerably flatter spectral slope. Finally, beyond a
	second crossover scale $k_{\rm II}$, the spectra approach yet another power-law
	regime with a significantly steeper decay.  The existence of these multiple
	scaling ranges suggests that the bottleneck is not merely a local accumulation
	of energy near the dissipative cutoff, but rather reflects the coexistence of
	qualitatively different dynamical states across scales.
	
	The observed scaling ranges can be understood by considering the equilibrium
	properties associated with the conserved invariant of the Voigt model
	$E_\alpha$. At sufficiently small wavenumbers, $\alpha^2 k^2 \ll 1$, so that
	the invariant reduces approximately to the ordinary quadratic energy.  In this
	regime, $k\lesssim k_{\rm I}$, the dynamics is expected to be essentially unaffected
	by the Voigt regularisation, and Kolmogorov scaling naturally persists.
	
	The intermediate regime $k_{\rm I} \lesssim k \lesssim k_{\rm II}$ may be interpreted by
	considering equipartition of the ordinary quadratic energy. In such a state,
	$\langle |u_n|^2\rangle \simeq \mathrm{const.}$, which, using the shell-model
	definition of the energy spectrum $E(k_n) = \frac{\langle
		|u_n|^2\rangle}{k_n}$, implies $E(k_n)\sim C_{\rm I} k_n^{-1}$, with a constant $C_{\rm I}$.
	
	Remarkably, this prediction agrees well with the scaling observed between
	$k_{\rm I}$ and $k_{\rm II}$ in the shell-model spectra, as seen in the compensated
	plots $k_n^{5/3}E(k_n)\sim k_n^{2/3}$ of
	Fig.~\ref{fig:CompSpectra}. The intermediate regime is therefore consistent
	with a state in which the ordinary quadratic energy has become approximately
	equipartitioned among the shells.
	
	The emergence of such an equilibrium-like state may be understood
	qualitatively from the scale-dependent slowdown introduced by the Voigt
	regularisation. Although the Voigt contribution remains subdominant in this
	range, it progressively increases the nonlinear transfer time at smaller
	scales. As a consequence, energy can no longer be transported efficiently
	through shell space and begins to accumulate at intermediate wavenumbers.
	Over sufficiently long times, this accumulation allows the dynamics to
	explore a larger fraction of the accessible phase space and approach a state
	that is approximately consistent with equipartition of the ordinary quadratic
	energy. In this sense, the intermediate regime may be viewed as a partially
	thermalised state arising from the weakening of the turbulent cascade.
	
	The high-wavenumber regime admits a similar interpretation. Once $\alpha^2 k^2
	\gtrsim 1$, the conserved invariant becomes dominated by its Voigt
	contribution, $E_\alpha \simeq \alpha^2 \sum_k k^2 |u(k)|^2$.  Equipartition of
	this quantity implies $\langle |u_n|^2\rangle \sim C_{\rm II} k_n^{-2}$, which immediately
	yields $E(k_n)\sim C_{\rm II} k_n^{-3}$ with $C_{\rm II}$ setting the magnitude of the 
	spectrum. 
	
	\begin{figure}
		\includegraphics[width = \linewidth]{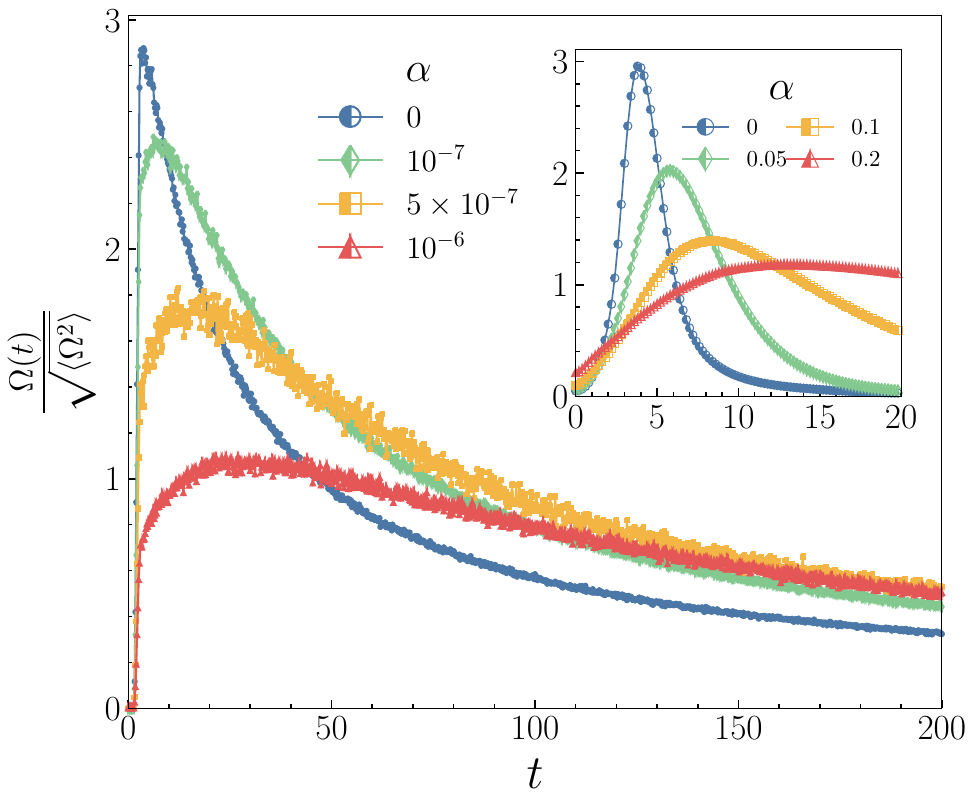} 
		\caption{Normalised enstrophy as a function of time for different values of the Voigt
			parameter $\alpha$ obtained from the shell model and (inset) 
			the corresponding results from our DNS of the Navier--Stokes--Voigt equations.
			Increasing $\alpha$ systematically shifts the enstrophy maximum to later
			times, indicating a progressive delay in cascade completion and providing
			evidence of the scale-dependent dynamical slowdown induced by the Voigt
			regularisation.}
		\label{fig:Omega}
	\end{figure}

	This prediction is in excellent agreement with the large-wavenumber scaling
	observed in Fig.~\ref{fig:CompSpectra}. The crossover between the two
	equilibrium regimes occurs when the two contributions to the invariant become
	comparable, $\alpha^2 k^2 \sim 1$, which defines the characteristic scale
	$k_{\rm II}\sim \alpha^{-1}$.
	
	The shell-model results therefore suggest that the bottleneck is not a single
	spectral feature but rather a composite structure consisting of two distinct
	equilibrium ranges associated with different asymptotic forms of the
	conserved Voigt invariant.
	
	Unlike the second crossover scale $k_{\rm II}$, whose location follows
	directly from the structure of the conserved invariant, the origin of
	the first crossover scale $k_{\rm I}$ is more involved. 
	Matching the spectra in Regimes II and III at
	$k_{II}\sim \alpha^{-1}$ implies that both may be
	parameterised by a common amplitude $\Theta$, which
	sets the equipartition level associated with the
	conserved invariant.
	Furthermore, a matching of the spectrum of Regimes I and 
	II at $k_{\rm I}$ yields the relation $k_{\rm I} \sim \varepsilon\Theta^{-3/2}$. 
	
	We can now estimate the total energy dissipation $\varepsilon$ in the
	shell model, Since the dissipation density grows with wavenumber,
	the contribution from the turbulent range $k \lesssim k_{\rm I}$ is
	asymptotically negligible, and the dominant contribution
	arises from the equilibrium-like ranges. Thus, 
	$\varepsilon = \nu \Theta \Sigma \frac{k_n^2}{1 + \alpha^2k_n^2}$ where
	the sum is over all wavenumber from $k_{\rm I}$ all the way up to the
	dissipative wavenumber $k_d \sim 1/\eta$, where $\eta = \left
	(\nicefrac{\nu^3}{\varepsilon}\right )^{1/4}$ is the Kolmogorov length
	scale. This allows us to obtain an explicit relation between $\Theta$
	and $\varepsilon$ and thence, up to log corrections 
	\begin{equation}
		k_{\rm I} = \eta^2k_{\rm II}^3. 
		\label{eq:kIshell}
	\end{equation}
	The crossover scale $k_{\rm I}$ therefore marks
	the point at which the forward cascade first becomes unable to sustain
	Kolmogorov-like transfer.  We refer the reader to Appendix~\hyperref[ap:derivation]{A} for a more
	rigorous derivation of this wavenumber and indeed the associated
	subleading terms.
	
	The broad inertial intervals accessible in the shell model make it
	possible to test this scaling prediction. In the inset of Fig.~\ref{fig:CompSpectra}, we
	test this prediction by rescaling the wavenumber axis with $k_I$ and
	plotting the compensated spectra. The transitions from the Kolmogorov
	range to the equilibrium-like regime collapse onto a single master
	curve for different values of $\alpha$, confirming the consistency of
	the estimate. To obtain a clean and reproducible determination of
	$k_I$, we employ a constant energy-injection forcing applied to the
	first two wavenumbers for this set of simulations.
	
	\begin{figure*}
		\includegraphics[width = \linewidth]{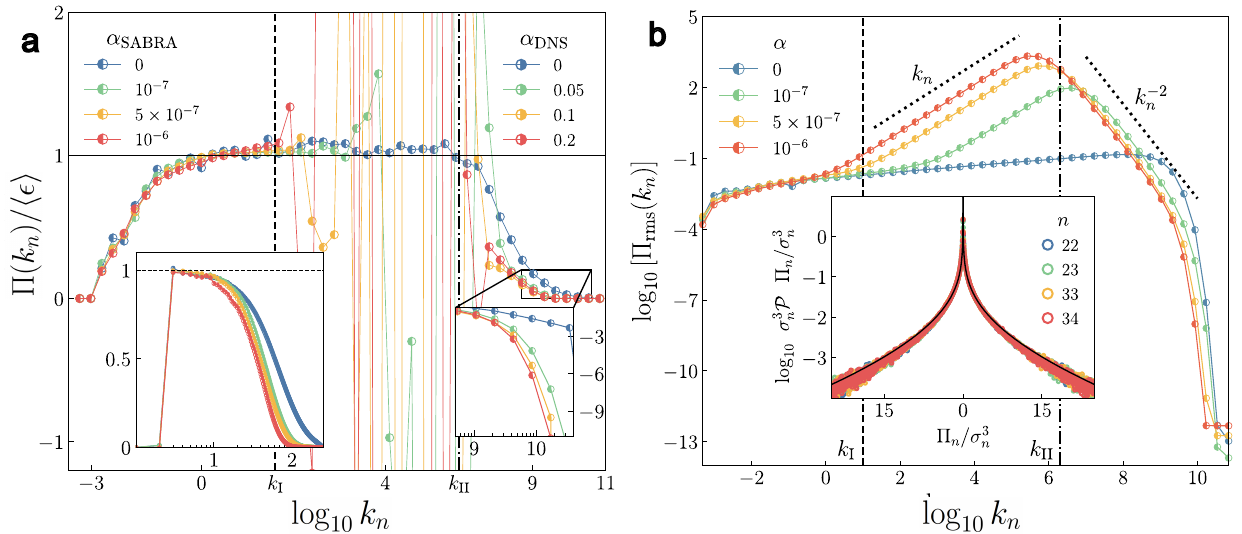}
		\caption{(a) Normalised energy Flux, $\Pi(k)/\varepsilon$, obtained from statistically
			stationary forced Voigt--SABRA simulations for different values of the Voigt
			parameter $\alpha$ (see legend). The vertical dashed and dash-dotted lines
			indicate the crossover wavenumbers $k_{\rm I}$ and $k_{\rm II}$, respectively, for the
			largest value of $\alpha$. The left inset shows the corresponding normalised
			flux obtained from DNS of the forced Navier--Stokes--Voigt equations,
			including $\alpha=0,$ the Navier--Stokes case. Both insets show the progressively faster decay of 
			the flux as the values of $\alpha$ are increased.
			(b) R.M.S of the energy flux averaged over time from forced Voigt--SABRA simulations, showing two scaling as predicted 
			from the theory (See text). The inset shows the p.d.f of the normalised energy flux for $\alpha=5\times 10^{-7}$ in the second
			and third regime following Eq.~\ref{eq:normal_cube}.
		}
		\label{fig:flux}
	\end{figure*}

	Additional insight is obtained from the decaying simulations shown in the
	right panel of Fig.~\ref{fig:CompSpectra}. In these runs, the spectra are
	measured at the instant when the enstrophy reaches its maximum value
	(see Fig.~\ref{fig:Omega}). This peak corresponds to the completion of the
	forward cascade and therefore provides a natural reference time for comparing
	simulations with different values of $\alpha$.
	
	A notable feature of the decaying simulations is that the time at which the
	enstrophy peak occurs increases systematically with increasing $\alpha$.
	The cascade therefore requires progressively longer times to reach the
	smallest scales, providing direct evidence of a slowdown of the turbulent
	dynamics induced by the Voigt regularisation. This behaviour is clearly
	visible in Fig.~\ref{fig:Omega}, where the enstrophy $\Omega$ is plotted as a
	function of time for both the shell model and DNS (inset).
	
	The decaying spectra provide particularly strong support for the equilibrium
	interpretation proposed above. In the absence of continuous energy injection,
	the equilibrium-like ranges become substantially more prominent. The
	shell-model spectra display clear signatures of both the intermediate
	$E(k_n)\sim k_n^{-1}$ regime and the large-wavenumber
	$E(k_n)\sim k_n^{-3}$ regime. The separation between the two equilibrium
	ranges is considerably easier to identify than in the forced simulations.
	This behaviour is consistent with the view that the intermediate and
	large-wavenumber regimes correspond to distinct equilibrium states associated
	with different asymptotic forms of the conserved invariant.
	
	Here it is important to distinguish the shell-model spectrum from the
	conventional three-dimensional energy spectrum. In the shell model, the
	spectrum is defined per logarithmic shell as $E(k_n)=\langle
	|u_n|^2\rangle/k_n$, whereas the three-dimensional energy spectrum
	includes the density of Fourier modes in a spherical shell. For
	three-dimensional turbulence, the equilibrium arguments discussed above
	therefore predict $E(k)\sim k^{d-1}=k^2$ and $E(k)\sim k^{d-3}=k^0$ for
	the intermediate $k_{\rm I} \lesssim k \lesssim k_{\rm II}$ and
	large-wavenumber $k \gtrsim k_{\rm II}$ equilibrium regimes,
	respectively.
	
	For the forced simulations, however, the DNS spectra remain dominated by the
	Kolmogorov-like regime. Increasing $\alpha$ leads to a systematic reduction
	of the inertial range together with an enhanced accumulation of energy at
	large wavenumbers. The available scale separation is insufficient to resolve
	the equilibrium ranges and consequently, the
	DNS spectra primarily exhibit the shrinking of the inertial range and the
	growth of the bottleneck but not Regimes II and III which show up clearly in the shell 
	model simulations.
	
	The decaying DNS spectra provide stronger support for the shell-model
	interpretation. When measured at the instant of maximum enstrophy, the
	spectra, see Fig.~\ref{fig:CompSpectra}(b) display a more pronounced tendency towards equilibrium-like behaviour
	at wavenumbers $k\gtrsim k_{\rm II}$. Although the available resolution
	remains insufficient to identify extended scaling ranges unambiguously, the
	observed trends are consistent with those found in the shell model.
	
	While the parameters of our decaying DNSs only reveal regime III, since
	$ k_{\rm II} \sim \mathcal {O}(1)$, it is nevertheless possible to
	carry out an analysis analogous to our shell model calculation to
	estimate $k_{\rm I}$ in the three-dimensional problem. By using the
	spectral matching technique as before (see also Appendix~\hyperref[ap:derivation]{A}), we
	estimate $k_{\rm I} = \eta^{-1}\left(\eta k_{\rm II}\right
	)^{6/11}$.
	
	While the spectral properties, summarised in Fig.~\ref{fig:CompSpectra}, provide a compelling, self-consistent 
	picture of the three regimes, when sufficiently scale-separated, the final evidence for this lies in the spectral properties 
	of the kinetic energy flux $\Pi$. In Fig.~\ref{fig:flux}(a) we show a plot of $\Pi (k_n)$, normalised by the mean dissipation rate $\varepsilon$, versus 
	the wavenumber $k_n$ from the shell model calculation; the pair of vertical lines denote $k_{\rm I}$ and $k_{\rm II}$. Clearly, in Regime I, the normalised 
	flux is unity and remarkably constant, consistent with the turbulent phenomenology and the spectral scaling in Fig.~\ref{fig:CompSpectra}. For wavenumbers,  
	$k_{\rm I} \lesssim k \lesssim k_{\rm II}$ the flux develops huge fluctuations and even alternate in signs from one wavenumber to another. This suggests a 
	non-definite flux and consistent with the equilibrium phenomenology. We can characterise this further from the probability density function (pdf) ${\mathcal P}$ 
	of the energy flux at a given shell number constructed from the time series of $\Pi (k_n)$. 
	Given the Gaussian nature of the shell velocities $u_n \sim \mathcal{N}(0,\sigma_n^2)$ 
	(Fig.~\ref{fig:gauss}), which is a consequence of equilibration, we approximate the flux 
	$\Pi (k_n) \sim k_n\, u_n u_{n+1} u_{n+2} \sim \big[\mathcal{N}(0,\sigma_n^2)\big]^3$ (by assuming weak 
	correlations between the velocities across shells) to obtain 
	\begin{equation}
		{\mathcal P} (\Pi (k_n)) = \frac{1}{3 \sqrt{2\pi \sigma_n^2}}\,|\Pi (k_n)|^{-2/3}\,
		\mathrm{exp}\left(-\frac{|\Pi (k_n)|^{2/3}}{2\sigma_n^2}\right).
		\label{eq:normal_cube}
	\end{equation}
	This is confirmed from our numerical simulations and a representative plot of the (normalised) ${\mathcal P}$, for four wavenumbers, is
	shown in the inset of Fig.~\ref{fig:flux}(b). Remarkably, the distributions for different wavenumbers agree very well with the theoretical 
	expression for ${\mathcal P}$, shown via a black line. Of course, the Eq.~\eqref{eq:normal_cube} is obtained from a strong assumption and hence the 
	lack of a perfect collapse in understandable. This pdf further strengthens the underlying
	physical picture: thermalisation is characterised
	not by the absence of energy exchange, but by the statistical cancellation of
	increasingly strong forward and backward transfers. From this perspective,
	$k_{\rm I}$ marks the scale at which the forward cascade weakens its directional
	dominance and equilibrium-like fluctuations begin to play a significant role.

	\begin{figure*}[t]
		\includegraphics[width = \linewidth]{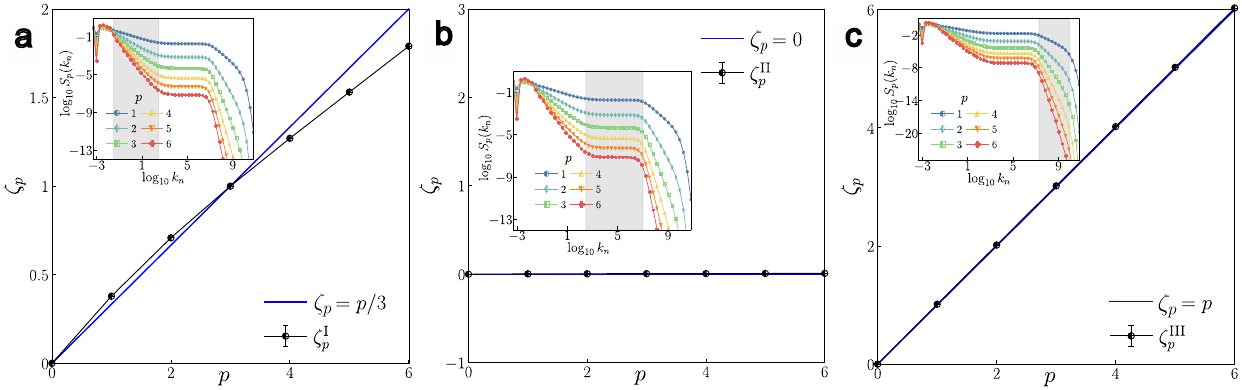}
		\caption{Equal-time scaling exponents obtained from the Voigt--SABRA shell model for
			$\alpha=10^{-7}$. Panels (a)--(c) show the exponents $\zeta_p$ as a function
			of order $p$ ($1\le p\le6$), evaluated in the low-wavenumber, intermediate,
			and high-wavenumber regimes, respectively. The corresponding equal-time
			structure functions $S_p$ are shown in the insets, with the fitting ranges
			indicated by the shaded regions. The three panels reveal distinct scaling
			behaviour in the turbulent, intermediate equilibrium-like, and Voigt-dominated
			equilibrium regimes identified from the energy spectra in
			Fig.~\ref{fig:CompSpectra}.}
		\label{fig:equal-time}
	\end{figure*}
	
	In the left inset of Fig.~\ref{fig:flux}(b) we show the shell-averaged flux from the DNS. The lack of scale separation leads not to the dramatic 
	fluctuations seen in the main panel for the shell model but a consistent weakening of the Kolmogorov range with increasing $\alpha$ and the onset 
	of the transition wavenumber $k_{\rm II}$. An exactly similar effect is seen in the right inset which zooms into the shell model flux near $k_{\rm II}$.
	
	Finally, we return once more to the fluctuations of the energy
	flux (Fig.~\ref{fig:flux}(a) and the associated inset in
	Fig.~\ref{fig:flux}(b)). Since this is seen only when the scale
	separation is significant --- as in our shell model
	calculations --- in what follows we restrict the analysis and
	notation to the SABRA model.  This
	picture can be quantified by considering the
	root-mean-square (RMS) flux. From the definition of the energy flux, 
	its RMS value can be approximated as $\Pi_{\mathrm{rms}}(k_n) \equiv \sqrt{\langle \Pi (k_n)^2 \rangle}
	\sim k_n \sqrt{\langle |u_n|^6 \rangle} \sim
	k_n^{1-\zeta_6/2}$.  Using the results of Sec.~IV, $\zeta_6=0$
	and $\zeta_6=6$ are expected in Regimes II and III,
	respectively, implying $\Pi_{\mathrm{rms}}(k_n)\sim k_n$ for
	$k_{\rm I} < k_n < k_{\rm II}$, and
	$\Pi_{\mathrm{rms}}(k_n)\sim k_n^{-2}$ for $k_n > k_{\rm II}$. 
	Our SABRA model data shows an exceptional agreement with this predicted scaling 
	form in Fig.~\ref{fig:flux}(b).

	The shell model therefore serves as the primary tool for resolving the detailed
	spectral structure of the bottleneck, while the DNS demonstrates that the same
	qualitative spectral evolution occurs in the underlying hydrodynamic equations.
	Taken together, these results suggest a picture in which three distinct
	spectral regimes coexist.  The existence of these three regimes raises the
	question of whether they correspond merely to spectral crossovers or to
	genuinely distinct statistical states. We address this issue in the next
	section.

	\section{Equal-time statistics and thermalisation}

	The spectral analysis of Sec.~III suggests the coexistence of three distinct
	dynamical regimes in Voigt-regularised turbulence: a turbulent inertial range
	at low wavenumbers, an intermediate equilibrium-like regime associated with
	partial thermalisation, and a high-wavenumber equilibrium regime dominated by
	the Voigt contribution to the conserved invariant. If this interpretation is
	correct, these regimes should also exhibit distinct signatures in their
	equal-time statistical properties. To investigate this question, we examine
	equal-time structure functions and probability distribution functions obtained
	from both shell-model simulations and DNS.
	
	The substantially larger scale separation available in the shell model allows
	reliable extraction of scaling exponents within each of the three spectral
	ranges identified in Fig.~\ref{fig:CompSpectra}. The corresponding equal-time
	structure functions are defined as
	\begin{equation}
		S_p(k_n)=\langle |u_n|^p\rangle \sim k_n^{-\zeta_p},
	\end{equation}
	where $\zeta_p$ denotes the scaling exponent of order $p$.
	
	Figure~\ref{fig:equal-time} shows the scaling exponents obtained independently
	within the three spectral ranges separated by the crossover scales $k_{\rm I}$ and
	$k_{\rm II}$. 
	
	The low-wavenumber regime, shown in Fig.~\ref{fig:equal-time}(a), displays the familiar
	nonlinear dependence of $\zeta_p$ on $p$ characteristic of intermittent
	turbulence in the SABRA model. The progressive suppression of intermittency
	by Voigt regularisation, observed previously in Ref.~\cite{Levant2010}, is
	here seen more clearly through the separation into the three spectral
	regimes. In particular, the anomalous scaling of the turbulent regime gives
	way to nearly simple scaling in the intermediate and high-wavenumber
	regimes.

	\begin{figure}
		\includegraphics[width = \linewidth]{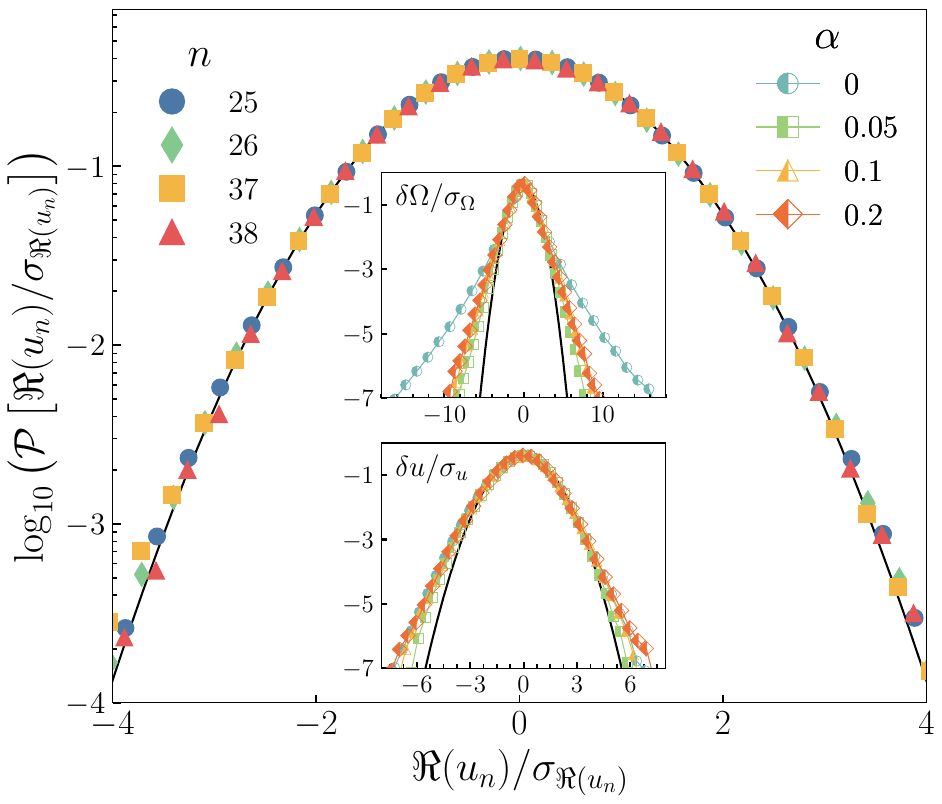}       
		\caption{Probability distribution function of
			the shell velocities $u_n$ obtained from Voigt--SABRA simulations for
			representative shells lying in the intermediate and high-wavenumber
			regimes identified in Fig.~\ref{fig:CompSpectra}. (Insets) The lower inset shows
			p.d.f. of velocity increments $\delta u_r$ obtained from DNS of the
			Navier--Stokes--Voigt equations, while the upper inset shows the
			corresponding p.d.f. of vorticity increments $\delta\Omega_r$. As the
			Voigt parameter $\alpha$ increases, the shell-model PDFs become
			progressively closer to Gaussian distributions. A similar trend is
			observed in the DNS results, particularly for the vorticity increments,
			although noticeable departures from Gaussianity persist.}
		\label{fig:gauss}
	\end{figure}
	
	\begin{figure*}[t]
		\includegraphics[width = \linewidth]{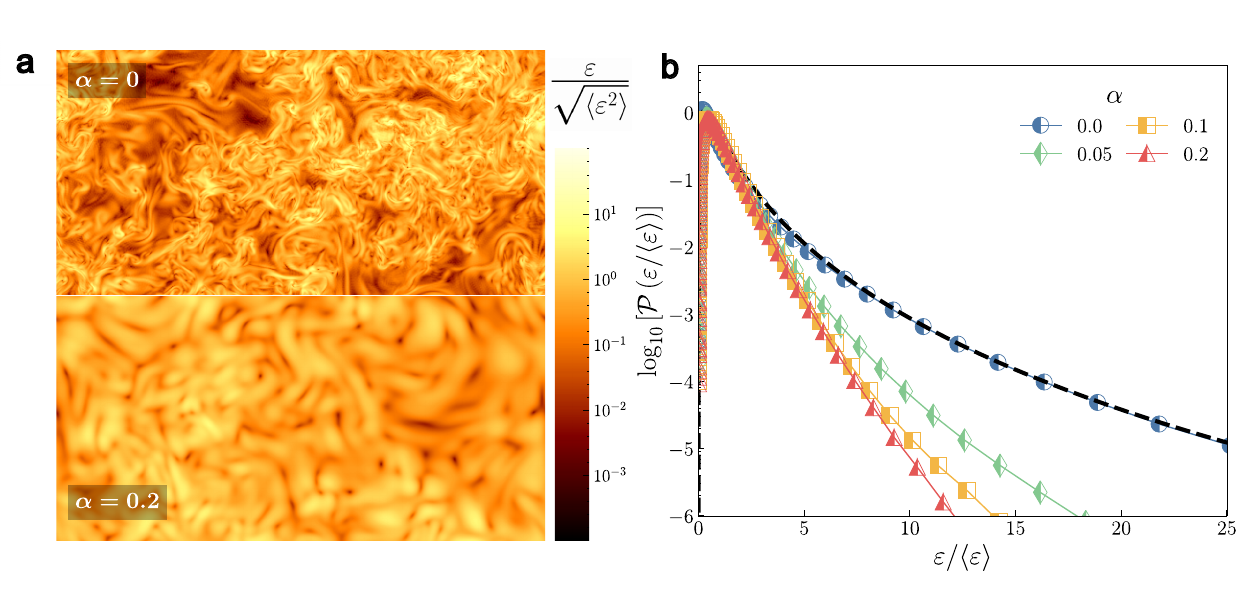} 
		\caption{Dissipation field and dissipation statistics from DNS of the
			Navier--Stokes--Voigt equations. (a) Pseudocolor plots of the
			dissipation field $\varepsilon = 2\nu S_{ij}S_{ij}$, where
			$S_{ij}=\frac{1}{2}(\partial_i u_j+\partial_j u_i)$ is the strain-rate
			tensor, for the classical Navier--Stokes case ($\alpha=0$) and a
			representative Voigt-regularised case ($\alpha=0.2$). The Voigt
			regularisation suppresses fine-scale filamentary structures and
			produces a smoother dissipation field. (b) PDFs of the normalised
			dissipation field $\varepsilon/\langle\varepsilon\rangle$ for different
			values of $\alpha$. Increasing $\alpha$ leads to a progressive
			narrowing of the distribution and a tendency towards Gaussian
			statistics, indicating reduced intermittency.}
		\label{fig:intermittency}
	\end{figure*}
	
	A qualitatively different behaviour emerges in the intermediate regime
	$k_{\rm I} \lesssim k_n \lesssim k_{\rm II}$ shown in
	Fig.~\ref{fig:equal-time}(b). Here the scaling exponents collapse towards
	$\zeta_p \simeq 0$, indicating that the structure functions become nearly
	independent of wavenumber. This behaviour is precisely what one expects if
	the shell amplitudes satisfy $\langle |u_n|^2\rangle \simeq \mathrm{const.}$,
	corresponding to equipartition of the ordinary quadratic energy. The
	structure-function analysis therefore provides independent support for the
	equilibrium interpretation inferred from the energy spectra.
	
	The high-wavenumber regime shown in Fig.~\ref{fig:equal-time}(c) exhibits yet
	another distinct scaling behaviour. In this range the exponents satisfy
	$\zeta_p \simeq p$, consistent with
	$\langle |u_n|^2\rangle \sim k_n^{-2}$, as expected from equipartition of the
	Voigt-dominated contribution to the conserved invariant. The structure-function
	exponents therefore provide strong evidence that the three spectral ranges
	identified in Sec.~III correspond to statistically distinct dynamical states
	rather than simple crossovers in the energy spectrum.
	
	The corresponding equal-time scaling exponents extracted from the fits are
	listed in Table~\ref{tab:zeta}, highlighting the transition from anomalous
	scaling in Regime I to simple scaling in Regimes II and III.
	
	Further evidence for this interpretation is obtained from probability
	distribution functions (PDFs), shown in Fig.~\ref{fig:gauss}. The main panel
	displays PDFs of shell-model velocity increments obtained from representative
	shells belonging to the intermediate and high-wavenumber equilibrium regimes.
	In both cases the distributions approach Gaussian statistics, in marked
	contrast to the strongly non-Gaussian behaviour characteristic of intermittent
	turbulence. Such behaviour is expected in thermalised systems and is
	consistent with the equilibrium interpretation suggested by both the spectral
	and structure-function analyses.
	
	The limited scale separation available in DNS prevents a similarly reliable
	determination of scaling exponents. In particular, the intermediate and
	Voigt-dominated ranges are not sufficiently extended in DNS to permit
	meaningful extraction of scaling exponents analogous to those obtained from the
	shell model. We therefore restrict the DNS analysis to probability
	distributions of physically relevant small-scale quantities. The inset of
	Fig.~\ref{fig:gauss} shows PDFs of vorticity increments $\delta\Omega_r$ for
	several values of the Voigt parameter.  The tails of the distributions are
	progressively suppressed with increasing $\alpha$, indicating a reduction in
	the probability of extreme dissipation events and, consequently, a suppression
	of small-scale intermittency.  The DNS therefore displays the same qualitative
	trend towards statistical equilibration observed in the shell model.
	
	Additional support is provided by the dissipation statistics shown in
	Fig.~\ref{fig:intermittency}. Panel (a) compares instantaneous
	dissipation fields obtained from DNS of the Navier--Stokes and
	Navier--Stokes--Voigt equations. Whereas the Navier--Stokes flow is
	characterised by intense filamentary structures associated with strong
	small-scale intermittency, increasing $\alpha$ progressively suppresses these
	fine-scale features and produces smoother, more spatially homogeneous
	dissipation patterns.
	
	This visual impression is quantified by the PDFs of the normalised
	dissipation field shown in Fig.~\ref{fig:intermittency}(b). As $\alpha$
	increases, the probability of extreme dissipation events decreases
	substantially and the distributions become progressively narrower. The Voigt
	regularisation therefore suppresses the strongest intermittent fluctuations
	and promotes a more homogeneous distribution of dissipation throughout the
	flow.
	
	\begin{table}
		\setlength{\tabcolsep}{6pt}
		\centering
		\begin{tabular}{|c|c|c|c|}
			\hline
			$p$ & Regime I & Regime II & Regime III \\
			\hline
			1 & 0.3789 $\pm$ 0.0001  & 0.0043 $\pm$ 0.0001 &  1.0134 $\pm$ 0.0001 \\
			2 & 0.7084 $\pm$ 0.0001 & 0.0057 $\pm$ 0.0004  & 2.0164 $\pm$ 0.0003  \\
			3 & 1.00 & 0.007 $\pm$ 0.001  & 3.0182 $\pm$ 0.0005  \\
			4 & 1.2699 $\pm$ 0.0001  & 0.007 $\pm$ 0.001  & 4.0207 $\pm$ 0.0009   \\
			5 & 1.5308 $\pm$ 0.0003 & 0.009 $\pm$ 0.002  & 5.0240 $\pm$ 0.0012  \\
			6 & 1.7897 $\pm$ 0.0004 & 0.010 $\pm$ 0.003 & 6.0277 $\pm$ 0.0015 \\
			\hline
		\end{tabular}
		\caption{Equal-time scaling exponents $\zeta_p$ measured in the three spectral regimes identified in 
			Fig.~\ref{fig:CompSpectra} from the shell model simulations for $\alpha = 10^{-7}$. 
			The exponents for Regime I were obtained via an Extended Self-Similarity (ESS) procedure.}
		\label{tab:zeta}
	\end{table}
	
	Taken together, the equal-time statistics provide strong support for the
	physical picture developed in Sec.~III. The low-wavenumber regime retains the
	intermittent characteristics of ordinary turbulence, whereas the intermediate
	and high-wavenumber ranges exhibit scaling exponents and probability
	distributions consistent with equilibrium-like behaviour. The DNS results,
	although limited by scale separation, display the same qualitative tendency
	towards Gaussian statistics and reduced intermittency, supporting the
	interpretation of the bottleneck as a manifestation of partial and eventual
	thermalisation induced by the Voigt regularisation.  Having established clear
	differences in the equal-time statistics of the three regimes, we next examine
	whether their temporal dynamics are similarly distinct.
	
	\section{Dynamic multiscaling and temporal equilibration}
	
	The spectral and equal-time statistics presented in Secs.~III and IV suggest
	the coexistence of three distinct dynamical regimes: a turbulent inertial
	range, an intermediate equilibrium-like regime, and a Voigt-dominated
	equilibrium regime. A natural question is whether these regimes can also be
	distinguished through their temporal dynamics. To address this issue, we
	examine time-dependent structure functions and the associated dynamic scaling
	exponents.
	
	Following Refs.~\cite{Mitra2004,Ray2008}, we define the time-dependent structure functions
	\begin{equation}
		F_p(k_n,t)= \frac{\left\langle \left[u_n(t_0)u_n^{*}(t_0+t)\right]^{p/2} \right\rangle} 
		{\left\langle |u_n(t_0)|^p \right\rangle},
	\end{equation}
	where the average is taken over different choices of the reference time
	$t_0$ in the statistically stationary state.
	
	From these functions we construct both integral and derivative time scales,
	\begin{equation}
		\mathcal{T}_{p,M}^{\mathrm{Int}}(k_n) = \left[ \int_0^{\infty} F_p(k_n,t)\,t^{M-1}\,dt \right]^{1/M},
	\end{equation}
	and
	\begin{equation}
		\mathcal{T}_{p,M}^{\mathrm{Der}}(k_n)=\left[\frac{\partial^M F_p(k_n,t)}{\partial t^M}\Bigg|_{t=0}\right]^{-1/M},
	\end{equation}
	which scale as $\mathcal{T}_{p,M}^{\mathrm{Int}}(k_n) \sim k_n^{-z_{p,M}^{\mathrm{Int}}}$ and $\mathcal{T}_{p,M}^{\mathrm{Der}}(k_n)
	\sim k_n^{-z_{p,M}^{\mathrm{Der}}}$.
	
\begin{figure*}
		\centering
		\includegraphics[width= \linewidth]{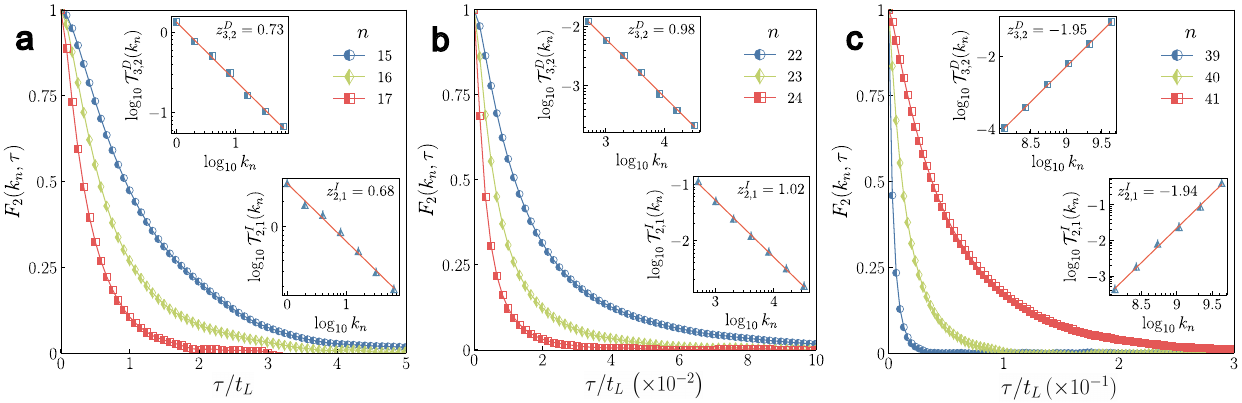}
		\caption{Second-order time-dependent structure functions for the three
			dynamical regimes identified in the Voigt--SABRA model for
			$\alpha=10^{-7}$. Panels (a)--(c) show the normalised structure
			functions $F_2(k_n,\tau)$ as a function of the dimensionless time
			increment $\tau/t_L$ for representative shells in the low-wavenumber
			(turbulent), intermediate (equilibrium-like), and high-wavenumber
			(Voigt-dominated equilibrium) regimes, respectively. The upper inset in
			each panel shows the derivative timescales $\mathcal{T}^{\mathrm{Der}}_{3,2}(k_n)$ and the 
			lower inset displays the corresponding integral timescales
			$\mathcal{T}^{\mathrm{Int}}_{2,1}(k_n)$ as a function of $k_n$ on log--log
			axes. The slopes of the least-squares fits yield the dynamic scaling
			exponents discussed in the text.}
		\label{fig:Fp}
	\end{figure*}

	Figure~\ref{fig:Fp} shows representative time-dependent structure functions plotted against 
	the dimensionless time increment $\tau/t_L$ and the corresponding dynamic exponents obtained independently within the
	three spectral ranges identified in Sec.~III. Here, 
	\begin{equation}
		\begin{aligned}
			t_L&=\displaystyle\sfrac{\left(\frac{\sum_n |u_n|^2/k_n^2}{\sum_n |u_n|^2/k_n}\right)}{\sqrt{\frac{2\sum_n |u_n|^2/k_n}{2\pi k_0}}},
		\end{aligned}
	\end{equation}
	denotes the box-size eddy turnover time. As in the equal-time analysis,
	the crossover scales $k_{\rm I}$ and $k_{\rm II}$ separate qualitatively distinct
	behaviours.
	
	The low-wavenumber regime, shown in Fig.~\ref{fig:Fp}(a), corresponds
	to the turbulent inertial range.  The measured dynamic exponents
	exhibit a nontrivial dependence on the order $p$, reflecting the
	presence of dynamic multiscaling. This behaviour is analogous to the
	anomalous scaling observed in the equal-time exponents and is
	consistent with the standard phenomenology of shell-model turbulence.
	The measured values remain close to the classical inertial-range
	estimate obtained from Kolmogorov theory, although systematic
	deviations are evident, particularly at higher orders.
	
	A qualitatively different picture emerges in the intermediate regime
	$k_{\rm I} \lesssim k_n \lesssim k_{\rm II}$ shown in
	Fig.~\ref{fig:Fp}(b). In Secs.~III and IV we argued that this range is
	consistent with an equilibrium-like state characterised by equipartition of
	the ordinary quadratic energy. The equal-time statistics exhibit simple
	scaling in this regime, suggesting that the temporal dynamics should likewise
	be governed by a single characteristic timescale.
	
	\begin{table*}
		\setlength{\tabcolsep}{10pt}
		\centering
		\begin{tabular}{|c|c|c|c|c|c|c|}
			\hline
			& \multicolumn{3}{c|}{$z_{p,1}^{\mathrm{Int}}$} & \multicolumn{3}{c|}{$z_{p,2}^{\mathrm{Der}}$} \\
			\hline
			$p$ & Regime I & Regime II & Regime III & Regime I & Regime II & Regime III \\
			\hline
			1 & 0.62 $\pm$ 0.02  & 0.99 $\pm$ 0.01 & -1.93 $\pm$ 0.04  & 0.69 $\pm$ 0.02  & 1.01 $\pm$ 0.02  & -1.95 $\pm$ 0.04 \\
			2 & 0.68 $\pm$ 0.02  & 1.02 $\pm$ 0.02  & -1.92 $\pm$ 0.03  & 0.72 $\pm$ 0.02  & 0.98 $\pm$ 0.01  & -1.93 $\pm$ 0.04 \\
			3 & 0.71 $\pm$ 0.01  & 1.02 $\pm$ 0.02  & -1.95 $\pm$ 0.02  & 0.73 $\pm$ 0.01  & 0.98 $\pm$ 0.01  & -1.92 $\pm$ 0.02 \\
			4 & 0.73 $\pm$ 0.01  & 1.01 $\pm$ 0.02  & -1.89 $\pm$ 0.03 & 0.74 $\pm$ 0.01 & 1.03 $\pm$ 0.01  & -1.90 $\pm$ 0.03 \\
			5 & 0.751 $\pm$ 0.009  & 1.00 $\pm$ 0.02  & -1.89 $\pm$ 0.04  & 0.75 $\pm$ 0.01 & 1.02 $\pm$ 0.01  & -1.91 $\pm$ 0.03 \\
			6 & 0.78 $\pm$ 0.01 & 1.02 $\pm$ 0.02  & -1.91 $\pm$ 0.03 & 0.75 $\pm$ 0.01  & 1.04 $\pm$ 0.01  & -1.92 $\pm$ 0.02\\
			\hline
		\end{tabular}
		\caption{Measured dynamic scaling exponents obtained from integral and derivative time scales in the three dynamical regimes from our shell model calculations for $\alpha = 10^{-7}$. The dimensional predictions are $z^{\rm I}\approx 2/3$, $z^{\rm II}=1$, and $z^{\rm III}=-2$.}
		\label{tab:dynamic}
	\end{table*}
	
	This expectation follows naturally from the modified nonlinear timescale
	introduced by the Voigt regularisation, $\tau_{\mathrm{nl}}^{-1}(k_n)$.  Since
	equipartition implies $\langle |u_n|^2\rangle \simeq \mathrm{const.}$, one has
	$|u_n|\sim k_n^0$.  Furthermore, throughout this regime $\alpha^2k_n^2 \ll 1$,
	so that $\tau_{\mathrm{nl}}(k_n) \sim k_n^{-1}$.  This dimensional argument
	therefore predicts the simple dynamic exponent $z^{\rm II}=1$.
	
	The measured integral and derivative dynamic exponents cluster around this
	value and exhibit negligible dependence on the order $p$. In contrast to the
	turbulent regime, dynamic multiscaling is therefore strongly suppressed in
	the intermediate equilibrium range.
	
	The high-wavenumber regime shown in Fig.~\ref{fig:Fp}(c) admits a similar
	interpretation. In this range, the equal-time statistics are consistent with
	equipartition of the Voigt-dominated contribution to the conserved invariant,
	implying $\langle |u_n|^2\rangle \sim k_n^{-2}$ and hence $|u_n|\sim k_n^{-1}$.
	Since $\alpha^2k_n^2\gg1$, the denominator in the nonlinear timescale is
	dominated by the Voigt term, yielding $\tau_{\mathrm{nl}}(k_n) \sim k_n^{2}$,
	which corresponds to the dynamic exponent $z^{\rm III}=-2$.
	
	Once again, the measured integral and derivative exponents are found to be
	close to this prediction and display no significant order dependence. The
	Voigt-dominated equilibrium regime is therefore characterised by a single
	dominant relaxation timescale rather than the hierarchy of timescales
	associated with turbulent multiscaling.
	
	For completeness, we note that the turbulent regime is expected to obey the
	usual bridge relations connecting equal-time and dynamic exponents. Such
	relations arise naturally from multifractal descriptions of turbulence and
	are broadly consistent with the measured exponents in
	Fig.~\ref{fig:Fp}(a). No analogous multifractal framework exists for the
	equilibrium regimes considered here. Instead, the observed exponents in
	Regimes II and III are accurately described by the simple dimensional
	arguments given above.
	
	Taken together, these results demonstrate that the transition from the
	turbulent regime to the two equilibrium regimes is reflected not only
	in the energy spectrum and equal-time statistics, but also in the
	temporal dynamics.  The turbulent regime exhibits dynamic multiscaling
	and a hierarchy of timescales, whereas the equilibrium regimes display
	simple temporal scaling governed by a single characteristic relaxation
	time. The dynamic exponents therefore provide further evidence that the
	bottleneck is associated with a progressive transition towards
	equilibrium-like small-scale dynamics induced by the Voigt
	regularisation.
	
	The measured integral and derivative dynamic scaling exponents are summarised
	in Table~\ref{tab:dynamic}. These results support the transition from dynamic
	multiscaling in the turbulent regime to simple temporal scaling in the
	intermediate and Voigt-dominated equilibrium regimes.

	\section{Conclusion}
	
	We have investigated the effects of Voigt regularisation on turbulent energy
	transfer using a combination of Voigt--SABRA shell-model simulations and direct
	numerical simulations of the Navier--Stokes--Voigt equations. Our results show
	that the primary effect of the Voigt term is not enhanced dissipation, but
	rather a scale-dependent slowdown of the nonlinear dynamics. As the Voigt
	parameter is increased, the forward cascade becomes progressively less
	efficient, leading to the formation of a bottleneck and the emergence of
	equilibrium-like behaviour at small scales.
	
	The shell-model simulations reveal that the bottleneck is not a single spectral
	feature but is instead associated with the coexistence of three distinct
	scaling regimes. At the largest scales, the dynamics remains turbulent and
	exhibits Kolmogorov-like scaling. At intermediate scales, the spectra,
	equal-time structure functions, and probability distribution functions are
	consistent with a partially thermalised state associated with equipartition of
	the ordinary quadratic energy. At sufficiently large wavenumbers, the dynamics
	crosses over into a second equilibrium regime associated with equipartition of
	the Voigt-dominated contribution to the conserved invariant.
	
	A simple interpretation of these regimes emerges from the structure of the
	conserved Voigt energy and the modified nonlinear timescale introduced by the
	regularisation. The crossover to the high-wavenumber equilibrium regime occurs
	at a scale $k_{\rm II}\sim \alpha^{-1}$, determined by the point at which the two
	contributions to the invariant become comparable. In contrast, the origin of
	the first crossover scale $k_{\rm I}$, which marks the breakdown of the turbulent
	cascade and the onset of partial thermalisation, is more subtle. 
	
	The equal-time and dynamic statistics provide further support for the
	thermalisation picture. In the turbulent regime, both equal-time and
	time-dependent structure functions exhibit the familiar signatures of
	multiscaling. In contrast, the intermediate and Voigt-dominated equilibrium
	regimes display simple scaling behaviour consistent with equipartition
	arguments. The dynamic exponents in these regimes are accurately described by
	dimensional estimates based on the modified nonlinear timescale, indicating
	that the temporal dynamics becomes progressively simpler as thermalisation sets
	in.
	
	The measured scaling exponents are broadly consistent with the theoretical
	expectations developed from equipartition arguments and the modified nonlinear
	timescale. Quantitative deviations remain (see Tables I and II), largely due to
	finite scaling ranges which affect exponent estimates, but the overall
	phenomenology and the separation into three distinct regimes are robust.
	
	Although the limited scale separation available in DNS prevents a
	quantitative identification of all scaling regimes, the DNS results
	display the same qualitative trends observed in the shell model,
	including bottleneck formation, delayed cascade completion, suppression
	of intermittency, and a tendency towards Gaussian statistics at small
	scales. The shell model therefore provides a detailed resolution of the
	three-regime phenomenology, while the DNS demonstrates that the
	associated bottleneck, dynamical slowdown, and suppression of
	intermittency persist in the underlying hydrodynamic equations.
	
	Taken together, these results support a view of bottleneck formation as
	a consequence of scale-dependent dynamical slowdown and the onset of
	equilibrium-like small-scale dynamics, rather than a purely dissipative
	effect.  More broadly, they suggest that Voigt regularisation provides
	a particularly transparent setting in which to study the competition
	between turbulent energy transfer and equilibrium dynamics, and may
	offer useful insights into the emergence of bottlenecks in more general
	turbulent systems.

	The present results also suggest a broader interpretation of bottleneck
	phenomena. In the Voigt system, the parameter $\alpha$ introduces
	additional crossover scales and effectively modifies the location at
	which nonlinear transfer becomes inefficient. Rather than being purely
	dissipative effects, bottlenecks may therefore arise whenever
	additional physical mechanisms renormalise the scales governing the
	cascade, leading to a partial accumulation and equilibration of energy
	before the ultimate dissipative cutoff is reached. The Voigt
	regularisation provides one explicit realization of this scenario
	through a scale-dependent slowdown of nonlinear interactions.  Whether
	analogous mechanisms operate in other regularised, multiscale, or
	geophysical turbulent systems remains an interesting direction for
	future work. In conclusion, in the regime $\alpha \ll \eta$, the
	Navier--Stokes--Voigt model reproduces the turbulent inertial range and
	turbulence statistics of the Navier–Stokes equations with high
	fidelity. This provides strong evidence that the Voigt model
	constitutes an excellent approximation to the Navier–Stokes equations
	for sufficiently small values of $\alpha$. From a mathematical
	perspective, however, the principal advantage of the
	Navier--Stokes--Voigt model is its global well-posedness for arbitrary
	initial data, including in the inviscid limit. In contrast, global
	regularity and well-posedness for the three-dimensional Navier–Stokes
	and Euler equations remain major open problems.
	
	\begin{acknowledgements}
		AK acknowledges Google Colaboratory where the DNS data was produced and analyzed. BP acknowledges the hospitality of ICTS-TIFR, Bengaluru and the support of the NIUS program of HBCSE-TIFR funded
		by the Department of Atomic Energy, Govt. of India (Project No. RTI4001).
		EST acknowledges the support in part by the DFG Research Unit FOR 5528 on Geophysical Flows.
		SSR acknowledges the Indo–French Centre for the
		Promotion of Advanced Scientific Research (IFCPAR/CEFIPRA, project no. 6704-1)
		for support.  This research was supported in part by the International Centre
		for Theoretical Sciences (ICTS) for the program --- 11th Indian Statistical
		Physics Community Meeting (code: ICTS/11thISPCM2026/04). The simulations were
		performed on the ICTS clusters Mario, Tetris, and Contra. AK and SSR
		acknowledge the support of the DAE, Government of India, under projects nos.
		12-R\&D-TFR-5.10-1100 and RTI4001. 
	\end{acknowledgements}

	\bibliographystyle{apsrev4-2}
	\bibliography{references}
	
	\appendix
	\onecolumngrid
	\section{Appendix A: Derivation of $k_{\rm I}$}
	\label{ap:derivation}
	\renewcommand{\theequation}{A-\arabic{equation}}
	\setcounter{equation}{0}  
	
	In this section we derive estimate of $k_{\rm I}$ with logarithmic corrections. To start with, we recall 
	\begin{equation}
		|u_k|^2 = \displaystyle
		\begin{cases}
			\displaystyle C_K\varepsilon^{2/3}k^{-2/3},\,\,\,&  k \lesssim k_{\rm I}\\
			\displaystyle \Theta,\,\,\,& k_{\rm I} \lesssim k \lesssim k_{\rm II}\\
			\displaystyle \frac{\Theta}{\alpha^2k^2},\,\,\,&  k_{\rm II} \lesssim k \lesssim k_{d}.
		\end{cases}
		\label{eq:relations}
	\end{equation} 
	where regime I and II are separated by $k_{\rm I}$ and regime II and III are separated by by $k_{\rm II} = \alpha^{-1}$; 
	the expression of $|u_k|^2$ in the third regime is obtained by equating the thermalised spectra with Voigt-modified spectra $(1+\alpha^2k^2)|u_k|^2=\Theta$.

	\subsubsection{Asymptotic dependence of $k_d$ on $\alpha$} Constantin,
	Levant and Titi \cite{CONSTANTIN2006120} showed that once the forcing
	is analytic (or finite-shell forcing), the shell amplitudes belong to a
	\textit{Gevrey} class $|u_k|\lesssim k^{-m}e^{-\sigma k^p}$ for some
	$m\ge0$, $\sigma>0$ and
	$p\leq\log_2(\nicefrac{1+\sqrt5}{2})\approx0.7$. Motivated by this
	result, we assume
	\begin{equation}
		|u_k|^2\sim\,k^{-2m}e^{-2\sigma k^p}.
	\end{equation}
	
\begin{figure*}[h]
		\centering
		\includegraphics[width=\linewidth]{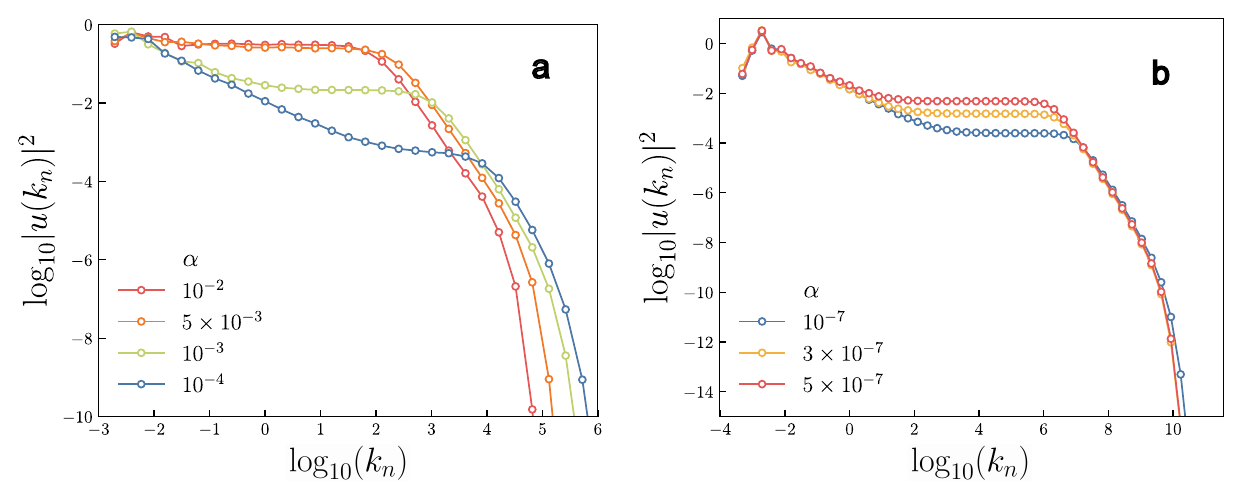}
		\caption{Log-Log plot of $|u(k_n)|^2$, as a function of $k_n$ for different values of the Voigt parameter $\alpha$ for (a)
			moderately large Reynolds number ($\nu = 10^{-8}$) and (b) extremely large Reynolds number ($\nu = 10^{-8}$).  
			For the case of (a) moderate Reynolds number, with $27.6\lesssim\ln(1/\nu\alpha^2)\lesssim36.8$, 
			the dissipation cutoff retains a visible $\alpha$-dependence, corresponding to a finite local exponent $\gamma_\alpha>0$. (This set 
			of simulations were performed with 36 shells and $k_0=1/512$; the forcing was restricted to $n=2$ and $3$ giving a mean energy injection 
			rate $\varepsilon=10^{-3}$.) 
			For much higher Reynolds number (panel (b)), the dissipation tails for different $\alpha$ (with $61.3\lesssim\ln(1/\nu\alpha^2)\lesssim64.5$) 
			collapse onto one another and 
			hence the dissipation wavenumber becomes independent of $\alpha$ as $\gamma_\alpha\rightarrow0$ in agreement with the theoretical prediction. 
			These simulations were performed with $N=48$, $k_0=1/2048$ and, as before, the forcing applied to shells $n=2$ and $3$ with $\varepsilon=10^{-3}$.} 
		\label{fig:k_d_verify}
	\end{figure*}

	To understand how the dissipation scale $k_d$ gets normalised by the Voigt term, we compare timescales of regime III and the regime where $k>k_d$:
	\begin{equation}
		\tau_{\rm nl}\sim \frac{\alpha^2 k_d}{u_k}\sim \frac{1}{\nu k_d^2}\sim \tau_\eta.
	\end{equation}
	By using the Gevrey estimate we have $k_d^{-2m}e^{-2\sigma k_d^p}\sim \nu^2\alpha^4k_d^6$ which yields, 
	\begin{equation}
		\sigma k_d^p+(3+m)\ln k_d\sim \ln\left(1/\nu\alpha^2\right).
	\end{equation}
	Let $y:=\ln\left(1/\nu\alpha^2\right)$, $A:=3+m$, $x:=k_d^p$ so that the implicit equation for the dissipation wavenumber becomes $\sigma x+\frac{A}{p}\ln x\sim y,$
	whose solution can be written in terms of Lambert's $W$ function
	\begin{equation}
		k_d\sim\left[\frac{A}{\sigma p}W(z)\right]^{1/p},
		\label{eq:kd_exact}
	\end{equation}
	where $z=\frac{\sigma p}{A}\exp\left(py/A\right)$. Let $k_d\sim \alpha ^{-\gamma_\alpha}$ then the local scaling exponent is defined by
	\begin{equation}
		\gamma_\alpha=-\left.\frac{\partial\ln k_d}{\partial\ln\alpha}\right|_\nu=\frac{2}{A\left(1+W\!\left(z\right)\right)}.
		\label{eq:gamma_exact}
	\end{equation}
	Taking the limit $y\gg 1$ or equivalently $\nu\alpha^2\ll 1$ and hence $z\to \infty$, we expand $W(z)$
	\begin{equation}
		W(z)= \ln z - \ln \ln z +\frac{\ln \ln z}{\ln z}+\cdots.
	\end{equation}
	Substituting this in Eq.~\ref{eq:gamma_exact} we obtain
	\begin{equation}
		\gamma_\alpha=\frac{2}{p\ln(1/\nu\alpha^2)}+\frac{2(3+m)}{p^2}\frac{\ln\ln(1/\nu\alpha^2)-1-\ln\sigma}{\ln^2(1/\nu\alpha^2)}+\cdots.
	\end{equation}
	Notice that in the limit $\nu\alpha^2\ll 1$, which is equivalent to the high Reynolds number limit, $\gamma_\alpha \to 0$. 
	Hence $k_d$ is independent of $\alpha$. Figure~\ref{fig:k_d_verify} provides compelling numerical support for this prediction. 
	While for moderately large Reynolds number flows, as shown in panel (a), the dissipation wavenumber clearly depends on $\alpha$, when the 
	Reynolds numbers are made substantially higher, as shown in Fig.~\ref{fig:k_d_verify}(b), $k_d$ is seen to be $\alpha$ independent.

	\subsubsection{Estimate of $k_{\rm I}$ in limit $\alpha^2\nu\ll 1$ for the Voigt shell model}
	
	The wavenumber $k_{\rm I}$ marks the scale for the transition from a turbulent spectra $E(k)=C_K\varepsilon^{2/3}k^{-5/3}$ to the equipartition one 
	$E(k)=\nicefrac{\Theta}{k}$. 
	Therefore, $k_{\rm I}\sim \varepsilon \Theta^{-3/2}$ and for a closed form expression for this wavenumber, as estimate of 
	$\Theta$.
	
	Consider
	\begin{align}
		\varepsilon
		&=\nu\sum_{k\lesssim k_{\rm I}} k^2|u_k|^2+\nu\sum_{k_{\rm I}\lesssim k \lesssim k_{\rm II}} k^2|u_k|^2+\nu\sum_{k_{\rm II}\lesssim k \lesssim k_d} k^2|u_k|^2+\nu\sum_{k> k_d} k^2|u_k|^2.
		\label{eq:posi}
	\end{align}
	Notice that the first term in the right hand side which denotes the dissipation in inertial turbulent regime can be neglected as 
	compared to the dissipation in the high wavenumbers. 
	
	We assume $|u_k|^2= B\,k^{-2m}e^{-2\sigma k^p}$. Matching this spectrum with the regime III spectrum at $k=k_d$ gives $\frac{\Theta}{\alpha^2k_d^2}=B k_d^{-2m}e^{-2\sigma k_d^p}$ with $B =\frac{\Theta}{\alpha^2}k_d^{2m-2}e^{2\sigma k_d^p}$. Hence,
	\begin{equation}
		|u_k|^2=\frac{\Theta}{\alpha^2}k_d^{2m-2}k^{-2m}e^{-2\sigma(k^p-k_d^p)}.
	\end{equation}
	We can similarly treat the high wavenumber dissipative term in Eq.~\ref{eq:posi}:
	\begin{align}
		\nu\sum_{k_n> k_d}k_n^2|u_{k_n}|^2=\frac{\nu\Theta}{\alpha^2}k_d^{2m-2}e^{2\sigma k_d^p}\sum_{k_n> k_d}k_n^{2-2m}e^{-2\sigma k_n^p}.
		\label{eq:TIV}
	\end{align}
	Let $k_d=k_{n_d}$. Writing $k_n=k_d\,2^j,$ $j=n-n_d>0,$ we obtain
	\begin{align}
		\frac{\nu\Theta}{\alpha^2}k_d^{2m-2}e^{2\sigma k_d^p}\sum_{j=1}^{\infty}(k_d2^j)^{2-2m}e^{-2\sigma(k_d2^j)^p}
		&=\frac{\nu\Theta}{\alpha^2}\sum_{j=1}^{\infty}2^{(2-2m)j}\exp\!\left[-2\sigma k_d^p(2^{pj}-1)\right]\nonumber\\
		&=
		\frac{\nu\Theta}{\alpha^2} \sum_{j=1}^{\infty}2^{(2-2m)j}e^{-\ell(2^{pj}-1)}= \frac{\nu\Theta}{\alpha^2} S(\ell),
	\end{align}
	where $\ell=2\sigma k_d^p$ is a dimensionless parameter. Now we estimate $S(\ell)$ for $\ell \gg 1$. For that, we define $a_j(\ell)=2^{(2-2m)j}e^{-\ell(2^{pj}-1)}$ and notice,
	\begin{equation}
		\frac{a_{j+1}(\ell)}{a_j(\ell)}=2^{2-2m}\exp\!\left[-\ell(2^p-1)2^{pj}\right]\to 0\qquad \text{ as $j\to\infty$.}
	\end{equation}
	Hence the series converges absolutely for every $m>0$. Moreover, notice that we can write, $a_j(\ell)=e^{-\ell(2^p-1)}2^{(2-2m)j}e^{-\ell(2^{pj}-2^p)},$ and the remaining series is uniformly convergent for every fixed $\ell>0$. Therefore,
	\begin{equation}
		S(\ell)=O\!\left(e^{-\ell(2^p-1)}\right)\to 0,\qquad \text{ as } \ell\rightarrow\infty.
	\end{equation}
	Now using the above result, we can write,
	\begin{align}
		\varepsilon
		&=\nu\sum_{k_{\rm I}\lesssim k \lesssim k_{\rm II}} k^2|u_k|^2+\nu\sum_{k_{\rm II}\lesssim k \lesssim k_d} k^2|u_k|^2=\nu\Theta\sum_{k_{\rm I}\lesssim k \lesssim k_{\rm II}}k^2+\frac{\nu\Theta}{\alpha^2}\sum_{k_{\rm II}\lesssim k \lesssim k_d}1 \nonumber\\
		&\sim \nu\Theta k_{\rm II}^2+\frac{\nu\Theta}{\alpha^2}\ln\left(\frac{k_d}{k_{\rm II}}\right)\sim \frac{\Theta\nu}{\alpha^2}\left\{1+\ln(\alpha k_d)\right\}
	\end{align}
	Therefore,
	\begin{equation}
		\varepsilon\sim \frac{\Theta\nu}{\alpha^2}\left\{1+\ln(\alpha k_d)\right\}=\frac{\nu\Theta}{\alpha^2}F(\alpha k_d)
		\label{eq:eps}
	\end{equation}
	We established that in the limit $\alpha^2\nu\ll 1$ or equivalently $Re\gg 1$, dissipation length scale is independent of $\alpha$ and hence, $F(\alpha k_d)\approx1+\ln(\alpha k_\eta):=F_\eta$ where $k_\eta=\left(\frac{\varepsilon}{\nu^3}\right)^{1/4}$ is the Kolmogorov scale. Hence we can eliminate $\Theta$ and obtain an estimate of $k_{\rm I}$ as
	\begin{equation}
		k_{\rm I}
		\sim\frac{\nu^{3/2}}{\sqrt{\varepsilon}\,\alpha^3}F_\eta^{3/2}
		\sim \eta^2k_{\rm II}^3F_\eta^{3/2}.
		\label{eq:k1}
	\end{equation}
	
	Similarly, let $k_{\rm I}\sim \alpha ^{-\Gamma_\alpha}$ then the local scaling exponent is defined by
	\begin{equation}
		\Gamma_\alpha
		:=-\frac{\partial\ln k_{\rm I}}{\partial\ln\alpha}\Bigg|_{\nu,\varepsilon}=3-\frac{3}{2}
		\frac{\partial\ln F_\eta}{\partial\ln\alpha}=3-\frac{3}{2F_\eta}.
	\end{equation}
	Hence the transition wavenumber possesses the asymptotic scaling $k_{\rm I}\propto\alpha^{-3},$ up to logarithmically varying corrections contained in $F_\eta$.
	
	It is important to make the following observation. Equation.~\ref{eq:eps} implies $\ln(\alpha k_d)>0$ and hence $\alpha>\eta$ for dissipation 
	to be non-negative. Therefore, we ought to have $\alpha>\eta$ for the additional Voigt regimes to appear. In particular, for $\alpha < \eta$, 
	the model is essentially indistinguishable from the $\alpha \to 0$ Navier-Stokes-Voigt equation as predicted earlier in Ref.~\cite{Levant2010}. 
	In principle, of course, choosing a sufficiently large $\alpha$ can lead to a vanishingly small turbulent regime as suggested in Fig.~\ref{fig:k_d_verify} 
	and also in the simulations of Ref.~\cite{Levant2010}.
	Thus, our work underlines the 
	fact that when $\alpha$ is much smaller than the dissipation length scale, the Voigt
	model leads to an inertial-range turbulent regime and
	turbulence statistics identical to the Navier–Stokes equations. Hence, these results  
	provide further evidence that the Voigt model constitutes an excellent
	practical approximation to the Navier–Stokes equations for small values
	of $\alpha$.
	
	\subsubsection{Estimate of $k_{\rm I}$ for the 3D Navier-Stokes-Voigt equation}
	
	For three-dimensional homogeneous isotropic turbulence, we define the energy
	spectrum through
	\begin{equation}
		\frac{1}{2}\langle |\mathbf{u}|^2\rangle=\int_0^\infty E(k)\,dk,
		\qquad
		\varepsilon=2\nu\int_0^\infty k^2E(k)\,dk,
	\end{equation}
	and the three regimes in Eq.~\ref{eq:relations}
	translate into
	\begin{equation}
		E(k)\sim
		\begin{cases}
			C_K\varepsilon^{2/3}k^{-5/3}, & k\lesssim k_{\rm I},\\
			\Theta k^2, & k_{\rm I}\lesssim k\lesssim k_{\rm II},\\
			\displaystyle\frac{\Theta}{\alpha^2}, & k_{\rm II}\lesssim k\lesssim k_d.
		\end{cases}
		\label{eq:3D_spectra}
	\end{equation}
	Therefore, one can perform a similar algebra as above and obtain $k_{\rm I}\sim\left(\frac{\varepsilon^{2/3}}{\Theta}\right)^{3/11}$.
	It remains to estimate $\Theta$, which again can be done by obtaining $\varepsilon$ as a function of $\Theta$:
	\begin{equation}
		\varepsilon\sim\frac{2\nu\Theta}{3\alpha^2}k_d^3,
	\end{equation}
	and hence $k_{\rm I}\sim\nu^{3/11}\varepsilon^{-1/11}k_d^{9/11}\alpha^{-6/11}$.
	In the limit $\nu\alpha^2\ll1$, $k_d$ becomes asymptotically
	independent of $\alpha$ and we obtain,
	\begin{equation}
		k_{\rm I}\sim\varepsilon^{5/44}\nu^{-15/44}\alpha^{-6/11}=\eta^{-1}\left(\eta k_{\rm II}\right)^{6/11}.
		\label{eq:kI_3D_final}
	\end{equation}
	Hence, defining the local scaling exponent by
	$k_{\rm I}\sim\alpha^{-\Gamma_\alpha}$, we find $\Gamma_\alpha=6/11$ in leading order.
	
\end{document}